\documentclass[12pt]{article}

\usepackage[utf8]{inputenc} 
\usepackage[T1]{fontenc}    
\usepackage{hyperref}       
\usepackage{url}            
\usepackage{booktabs}       
\usepackage{amsfonts}       
\usepackage{nicefrac}       
\usepackage{caption}
\usepackage{graphicx}
\usepackage{authblk}
\usepackage{titlesec}
\usepackage{natbib}
\usepackage{amsmath}
\usepackage{svg}
\usepackage{makecell}
\usepackage{multirow}
\usepackage{chngcntr}
\usepackage[title]{appendix}

\setcitestyle{authoryear,round,semicolon}

\usepackage{caption, setspace}
\DeclareCaptionFont{helv}{\normalsize\fontfamily{phv}\selectfont}

\usepackage{geometry}
\title{\normalsize
Efficient optimization of ODE neuron models using gradient descent
}

\author[1,2*]{Ilenna Simone Jones}
\author[2,3]{Konrad Paul Kording}
\affil[1]{Kempner Institute for the Study of Natural and Artificial Intelligence at Harvard University, Harvard University}
\affil[2]{Department of Neuroscience, University of Pennsylvania}
\affil[3]{Department of Bioengineering, University of Pennsylvania}
\affil[*]{Corresponding author: Ilenna Jones, ijones@fas.harvard.edu}

\begin{document}

\maketitle
\date{}
\begin{quote} 	{\normalfont\normalsize\bfseries\centering Abstract\par}

\indent
Neuroscientists fit morphologically and biophysically detailed neuron simulations to physiological data, often using evolutionary algorithms. However, such gradient-free approaches are computationally expensive, making convergence slow when neuron models have many parameters. Here we introduce a gradient-based algorithm using differentiable ODE solvers that scales well to high-dimensional problems. GPUs make parallel simulations fast and gradient calculations make optimization efficient. We verify the utility of our approach optimizing neuron models with active dendrites with heterogeneously distributed ion channel densities. We find that individually stimulating and recording all dendritic compartments makes such model parameters identifiable. Identification breaks down gracefully as fewer stimulation and recording sites are given. Differentiable neuron models, which should be added to popular neuron simulation packages, promise a new era of optimizable neuron models with many free parameters, a key feature of real neurons.
\end{quote}

\newpage

\section{Introduction}

Single neuron models are often used for generating and testing hypotheses, as well as building blocks to model multi-neuronal systems. Depending on the question at hand, modelers choose what degrees of detail or abstraction to impart on the neuron model \citep{frigg_models_2020, poirazi_illuminating_2020}. As experimental techniques provide higher resolution imaging at dendritic resolutions over wider fields of view \citep{ali_interpreting_2019, landau_dendritic_2022, aggarwal_glutamate_2023, cornejo_voltage_2021}, it becomes clearer that variability in neural data may require computational models with more degrees of freedom, or higher number of parameters. 
For example, classical models and reduced models assume that all channel densities are constant across the dendritic tree \citep{van_geit_bluepyopt_2016, gouwens_systematic_2018, amsalem_efficient_2020, wybo_data-driven_2021}, despite knowledge to the contrary of heterogeneous distributions that would naturally require many more parameters \citep{stuart_dendrites_2016, komendantov_dendritic_2009, migliore_signal_2005}.
The transition to large models thus requires the development of approaches, including parameter tuning, to fit such models to neuroscience data.


The process of fitting models to data requires tuning the model's parameters. The popular methods  are 0th order methods that require at least one iteration of forward simulation to calculate the goodness of fit and the necessary change per parameter of a model \citep{gegenfurtner_praxis_1992, van_geit_automated_2008, van_geit_bluepyopt_2016}. This is one dimension of feedback per calculated change in a parameter. Commonly used algorithms (e.g. evolutionary optimization) run the simulation, compare results to physiology, and iterate, making progress by proposing changes to parameters and rejecting bad parameter settings. Intuitively, we can directly see why these 0th order algorithms will have trouble if there are many parameters; we need to loop through all parameters to incrementally improve each of them once. Fitting large models with these methods is a difficult problem, often requiring large-scale compute and  motivating model reduction strategies \citep{van_geit_bluepyopt_2016, amsalem_efficient_2020}. 

Gradient descent is a 1st order method, which makes use of multidimensional gradients and not merely the one dimensional goodness of fit. While 0th order methods use only fit quality, and thus really learn about sensitivity only in one dimension, 1st order methods use one calculation for changing all parameters and learn about sensitivity to all parameters at the same time. This allows updating all parameters after running the simulation only once. For detailed models, which have high parameter counts this should allow significant speedups. Indeed, gradient descent is used efficiently in artificial neural networks (ANNs) to fit massive models with more than $10^{9}$ model parameters \citep{rumelhart_learning_1986, kaplan_scaling_2020, muennighoff_scaling_2023}. It is thus  attractive to use 1st order methods to optimize neuron models.

What makes neuron models different from ANNs is that they require integration of the relevant differential equations over time. Simulation of such models  are defined by systems of ordinary differential equations (ODEs). Their simulation thus requires numerical ODE solvers, which are usually not differentiable. However, developments \citep{chen_neural_2019} have introduced  efficient differentiable ODE solvers for the deep learning field. For example, there are now good ODE solvers available in PyTorch \citep{paszke_pytorch_2019}, a Deep Learning programming library. As such, it seems that neuron simulations are now compatible with 1st order optimization techniques, promising more efficient optimization. 

Here we demonstrate the use of differential ODE solver to efficiently optimize simulations of single neurons with heterogeneous ion channel distributions in active dendrites. Using gradient descent we optimize the parameters of multicompartmental neuron models, defined by cable equations with voltage-gated ion channels. Such models have many parameters (e.g., those characterizing the ion channel density). To do this, we implement biophysical neuron models in PyTorch so that the full forward operation of mapping neuron model input to output is differentiable. Using standard optimizers popular in the deep learning field we find rapid convergence. We verify on simulated neurons that we can invert the generative models and recover the true parameters when recording and stimulating all compartments and characterize how model parameter inference degrades as recording and stimulation of fewer compartments is used. 

\section{Methods}
\subsection{The general problem of optimization}

This paper explores the advantages of 1st order vs 0th order optimization in the context of detailed neuron models. We thus want to start building some intuitions about optimization and 1st vs 0th order methods. We will start discussing the general form of the involved optimization problems. Let us consider that we have the general form of a time varying model:
\begin{align}
{\bf y} = f({\bf x}, w)
\end{align}
where
\begin{quote}
    ${\bf x}=x(t)$ is a time-varying input vector
 
    $w$ is a parameter vector
 
    ${\bf y}=y(t)$ is an output vector
\end{quote}

\noindent By evaluating model $f$, parameterized by a set of parameters $w$, using input $\bf x$ we can find $\bf y$. You can then say that $f$ maps input $\bf x$ to $\bf y$ , or that this model describes the input-output relationship of the data $({\bf x}, {\bf y})$. For the kinds of models we are considering, $f$ is not directly defined but rather through an ODE:

\begin{align}
    \frac{dy}{dt}=\frac{d}{dt}f({\bf x}, w) =g({\bf x},w)
\end{align}

for a suitably defined $g$ which characterizes all our assumptions about the structure of our neuron model.
 
We can evaluate how well the model estimate of the output, ${\bf y}$, matches the target data, $\hat{\bf y}$. We calculate the model mismatch, or {\it loss}  ($\mathcal L$), using an error function (or objective function), $\mathcal L(w)$, where $\mathcal{L} \in \mathbb{R}$. Given we are trying to approximate the input-output relationship for $(\bf x, \hat{y})$, the error function changes as a function of the parameters, represented by parameter vector $w$. To approximate this input-output relationship using the model, we want to minimize model mismatch, $\mathcal L$, using an optimization of $w$. We can represent this relationship as $\mathcal{L}({\bf y, \hat{y}}) = \mathcal{L}(f({\bf x}, w), \hat{\bf y}) =  \ \mathcal{L}(w)$.

\subsection{The benefits of first order vs zeroth order optimization}

Optimization can be split into two cases: 1) Initialization of $w$ is far from the optimum, where the loss "landscape" can take any shape, and 2) local optimization where the loss function becomes roughly quadratic.

Let us consider the local quadratic approximation to the loss function close to the minimum, where the Taylor series expansion is precise. Let us assume for convenience that the minimum is at $\mathcal{L}=0$.

Consider the Taylor expansion of $\mathcal{L}(w)$ around some point $\hat{w}$ in weight space: 

\begin{align}
    \mathcal{L}(w) \approx \mathcal{L}(\hat{w}) + (w-\hat{w})^T \nabla \mathcal{L}(\hat{w}) + \frac{1}{2}(w-\hat{w})^T \nabla \nabla \mathcal{L}(\hat{w}) (w-\hat{w})
\end{align}
$\nabla \mathcal{L}(\hat{w})$ or $\nabla \mathcal{L}$ is the gradient of $\mathcal{L}$ evaluated at $\hat{w}$ and $\nabla \nabla \mathcal{L}(\hat{w}$ or $\nabla^2 \mathcal{L}$ is the Hessian (a square matrix of partial second derivatives) of $\mathcal{L}$ evaluated at $\hat{w}$. These correspond to 0th, 1st, and 2nd order terms that can be used to approximate $\mathcal{L}(w)$. An optimization algorithm's goal is to use any combination of these terms to approximate where $\mathcal{L}(w) = 0$.

Let's observe that:
\begin{quote}
    $w \in \mathbb{R}^W$
 
    $\nabla \mathcal{L} \in \mathbb{R}^W$
 
    $\nabla^2 \mathcal{L} \in \mathbb{R}^{W\times W}$
\end{quote}

\noindent We can use the fact that $\nabla^2 \mathcal{L}$ is symmetric to determine that the full $\mathcal{L}$ surface has $W(W+3)/2$ independent elements. Therefore the location of the minimum depends on $\mathcal{O}(W^2)$ values. We thus need $\mathcal{O}(W^2)$ independent pieces of information to locate the minimum \citep{bishop_pattern_2016}.

Let's consider using only 0th terms to find optimal $w$.  Let's say evaluating $\mathcal{L}(w)$ takes some constant $p$ amount of time. The 0th-order term,  $\mathcal{L}(\hat{w})$, when evaluated produces one piece of information that can be used to locate the minimum. To produce on the order of $W^2$ pieces of information, it should then take $\mathcal{O}(pW^2)$ amount of time. This means that 0th-order methods scale quadratically in time with respect to the number of parameters. This can be prohibitive with highly-parameterized models.

Let's consider using only 1st-order terms to find optimal $w$. Evaluating the gradient produces W pieces of information because $\nabla \mathcal{L} \in \mathbb{R}^W$. To evaluate $\nabla \mathcal{L}$, we could determine each element of $\nabla \mathcal{L}$ by calculating $\frac{\partial \mathcal{L}}{\partial \hat{w_i}}$ in series using methods such as finite differences, which takes constant time $q$ \citep{bishop_pattern_2016}. This would thus take $\mathcal{O}(qW)$ to produce $W$ pieces of information for each evaluation of $\nabla \mathcal{L}$, and it would thus take $\mathcal{O}(qW^2)$ time to produce $W^2$ pieces of information.

Machine learning field introduces the backpropagation of error algorithm to compute the gradient of the loss with respect to the model's parameters, thereby replacing finite differences. The "backprop" algorithm evaluates all elements of $\nabla \mathcal{L}$ in parallel, which we will say takes constant time $r$ \citep{rumelhart_learning_1986}. This means that it would take $\mathcal{O}(r)$ time to produce $W$ pieces of information for each evaluation of $\nabla \mathcal{L}$, and it would thus take $\mathcal{O}(rW)$ time to produce $W^2$ pieces of information. This means that first-order methods that rely on backpropagation scale linearly in time with respect to the number of parameters, making them generally more efficient for optimizing highly-parameterized models.

We can thus intuitively see that as the dimensionality of the problem gets large, 1st order methods scale much better than 0th order methods.


\subsection{Defining the neuron model}

\begin{figure}[!t]
  \centering
  \includegraphics[width=0.8\linewidth]{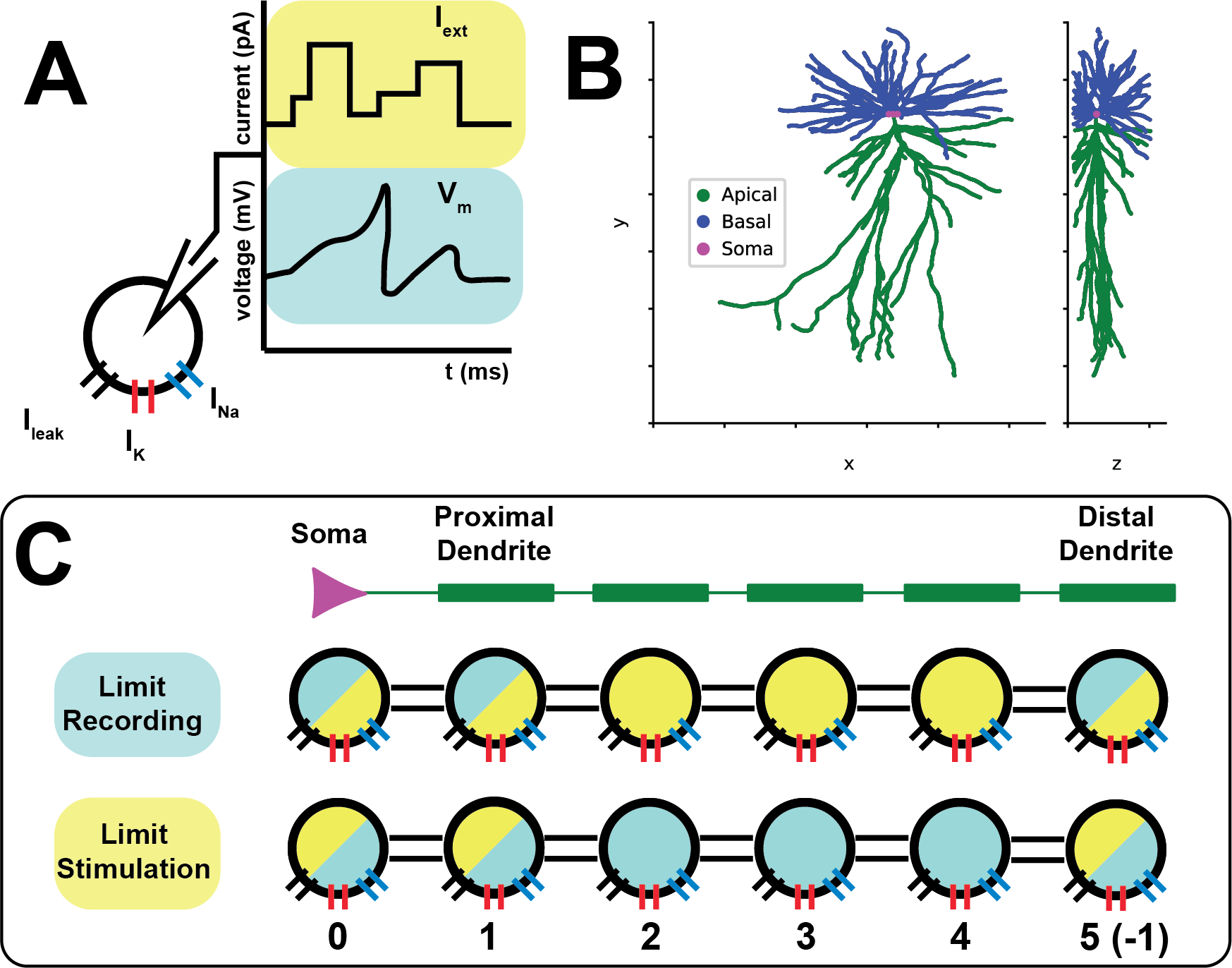}
  \caption{\textbf{Neuron models with dendrites and their parameters.} (a) We use a biophysical neuron model that has one optimizable parameter per voltage-gated ion channel current, scaling the magnitude of $I_{Na}$ and $I_K$.  $I_\text{leak}$ is not changed via optimization. A variable step current is used as the external input ($I_\text{ext}$), resulting in a membrane voltage trace ($V_m$). (b) The \textit{Test} model, which is an 184-compartment, mouse layer 2/3 pyramidal cell model. It has 2 parameters per compartment, resulting in 368 optimizable parameters. (c) The \textit{Toy} model  is a 6 compartment ball-and-stick model with 12 optimizable parameters.}
  \label{fig:model_comp}
\end{figure}

There are many options for neuron models, but for this paper we will use a simple biophysical conductance-based model (Fig.~\ref{fig:model_comp}A) :

\begin{align}
I_m &= I_{Na} + I_{K} + I_\text{leak} + I_\text{ext}
\end{align}
which expands to:
\begin{align}
C_m \frac{dV}{dt} &= \overline{g_{Na}}g_{Na}(V,t)(V - E_{Na}) + \overline{g_{K}}g_{K}(V,t)(V - E_{K}) + \overline{g_{leak}}(V - E_{leak}) + I_\text{ext}
\end{align}

where
\begin{quote}
    
    $C_m$ is the capacitance across the cell membrane
 
	$V = V(t)$ is the voltage difference across the cell membrane

        $I_\text{ext}$ is externally supplied current
        
	$\overline{g_{Na}}$ and $\overline{g_{K}}$ corresponds to maximal conductance, or "channel density"
 
	$g_{Na}(V,t) \in [0,1]$ and $g_{K}(V,t) \in [0,1]$ correspond $Na^+$ and $K^+$ channel opening kinetics
 
	$E_{Na}$, $E_{K}$, and $E_{leak}$ are the reversal potentials of each conductance.
    
\end{quote}

There are three types of parameters in such a model: the maximal conductance parameters $\overline{g_{Na}}$ and $\overline{g_{K}}$, the kinetics parameters that define the time-varying channel gating functions $g_{Na}(V,t)$ and $g_{K}(V,t)$, and the passive parameters $C_m, E_{Na}$, $E_{K}$, and $E_{leak}$ . To keep the identity of the channels whose kinetics are defined by the parameters in these channel gating functions, we do not optimize the kinetics parameters (they are arguably specified by evolution). We also do not optimize the passive parameters, since they are a property of the neuron's environment or the neuronal morphology. Instead, we optimize only the maximal conductance parameters $\overline{g_{Na}}$ and $\overline{g_{K}}$, which corresponds to how much of each type of channel conductance is present in the membrane of the neuron model.

We can expand on this definition of a model to make it multicompartmental by adding axial current terms following Kirchhoff's Current law:

\begin{align}
I_m^{(i)} &= I_{Na} + I_{K} + {I_{leak}} + I_{m}^{(i-1)} + I_{m}^{(i+1)} + I_\text{ext} \\
I_m^{(i)} &= I_{Na} + I_{K} + {I_{leak}} + g_{ax}^{(i+1)}(V^{(i+1)} - V^{(i)}) + g_{ax}^{(i-1)}(V^{(i-1)} - V^{(i)}) + I_\text{ext}
\label{eq:hh}
\end{align}

where
\begin{quote}
    $g_{ax}^{(i)}$ is the axial conductance connecting a compartment with voltage $V^{(i)}$ to the target compartment
\end{quote}

Axial conductance is a function of the diameter of the branch of dendrite. We assume that the axial conductance is a constant throughout the dendritic tree (See Table~\ref{tab:simparams}).

\subsection{A toy model and a more realistic neuron model}

In this demonstration of how gradient descent can be used to efficiently optimize differentiable neuron models, we define two multicompartment models for which we then optimize the maximal conductance parameters.  We implement a 6 compartment model which we call the toy model (Fig.~\ref{fig:model_comp}D). We also implement a 184 compartment, morphologically realistic model, based on the model of a cortical layer 2/3 pyramidal cell defined in \cite{gidon_dendritic_2020} and call that the test model (Fig.~\ref{fig:model_comp}C). In other methods \citep{van_geit_bluepyopt_2016, gouwens_systematic_2018}, to minimize the number of parameters to optimize the neuron model, the "active properties," which are the voltage gated ion channel conductances, were localized to the soma compartment only, leaving the dendrites to remain passive cable compartments. Models with active dendrites use constant maximal conductance parameters, assuming channel density is homogeneous so as to reduce model complexity \citep{van_geit_bluepyopt_2016, gouwens_systematic_2018, gidon_dendritic_2020}.  To demonstrate that gradient descent can be used to optimize biologically relevant degrees of freedom, we optimize these models as if the dendrites have active properties heterogeneously distributed in their dendritic trees. Since there are 2 maximum conductance parameters per compartment, therefore the toy model has 12 learnable parameters and the test model has 368 learnable parameters. Thus, we optimize the maximal conductance parameters for each compartment to infer how channel densities are distributed across the dendritic tree.

In the original \cite{gidon_dendritic_2020} model, the parameter values were $\overline{g_{Na}} = 100 \text{mS}$ and $\overline{g_{K}} = 45 \text{mS}$. We used these values as the initial parameters of the untrained model and then fit these parameters to target data generated by a ground truth target model. The target model's compartmental parameters were randomly sampled as defined by $\overline{g_{Na}} \sim \mathcal{U}(1, 0.3)*(100 \text{mS})$ and $\overline{g_{K}} \sim \mathcal{U}(1, 0.3)*(45 \text{mS})$. These randomized parameters were our "ground truth" parameter values $w$ that our optimization method had to then infer or approximate.

To produce the target data $\hat{V}$, we provided rich, varied current input $I_{ext}$ by randomizing step currents to each compartment within a range of 0 to 20 pA where the currents change 5\% of the total simulation time. 100 sets of randomized step current traces were provided to each compartment as input into each model, yielding 100 sets of voltage traces from each model (see Figure~\ref{fig:model_comp}A for an example).

\begin{table}[!t]
    \centering
    \small
    \begin{tabular}{|l|r|r|}
        \toprule
        \multicolumn{3}{c}{\textbf{Simulation Parameters}} \\
        \midrule

        {} &    Toy &    Test \\ \hline
        Compartments                  &   6 &  184 \\
        Axial Conductance ($\mu$S)      &   0.50 &    0.01 \\
        Minimum Stimulus Current (pA) &   0.00 &    0.00 \\
        Maximum Stimulus Current (pA) &  20.00 &   10.00 \\
        Simulation Duration (ms)      &   5.00 &    5.00 \\
        dt (ms)                       &   0.10 &    0.10 \\
        Stimulus Hazard Rate          &   0.05 &    0.05 \\
        \bottomrule
    \end{tabular}
    \caption{Simulation parameters for multicompartment neuron models.}
    \label{tab:simparams}
\end{table}

\subsection{Differentiable ODE solvers to optimize neuron models}

We demonstrate that a neuron model can not only be optimized using gradient descent, but that it is also possible to use the differentiable ODE solver to feasibly do so. We must then show we fulfill the necessary conditions for gradient descent.

Gradient descent is of the general form:
\begin{align}
w_{\tau+1} = w_{\tau} - \eta \nabla \mathcal{L}(w_{\tau})
\label{eq:gd}
\end{align} 
where
\begin{quote}
    $\eta$ is the learning rate
    
    $w_\tau$ is the current parameter vector
    
    $w_{\tau+1}$ is the next parameter vector after applying changes using the gradient information
\end{quote}

The parameter vector for our biophysical conductance parameter is:

\begin{align}
    w = \begin{bmatrix} \overline{g_{Na}} \\ \overline{g_{K}}  \end{bmatrix}
\end{align}
where each element is a row vector consisting of parameter values for each compartment. For example, $\overline{g_{Na}} = [\overline{g_{Na_1}} , \overline{g_{Na_2}} , ... , \overline{g_{Na_n}}]$ where $n$ is the number of compartments.

Then the gradient of the error with respect to the parameters must be:
\begin{align}
    \nabla \mathcal{L}(w) = \begin{bmatrix} \frac{\partial \mathcal{L}}{\partial \overline{g_{Na}}} \\ \frac{\partial \mathcal{L}}{\partial \overline{g_{K}}} \end{bmatrix}
\end{align}

We can approximate $\nabla \mathcal{L}(w)$ with the backprop algorithm, which is an efficient way to calculate the gradients of the error with respect to each parameter of the model. Backprop is an application of the chain rule in the following way:
\begin{align}
    \nabla \mathcal{L}(w)= \frac{\partial \mathcal{L}}{\partial w_i} = \frac{\partial \mathcal{L}}{\partial y_i} \frac{\partial y_i}{\partial w_i} 
\end{align}
where
\begin{quote}
    $w_i$ is a single parameter in the parameter vector of the model
    
    $y_i$ is the activation or state variable that results from using the parameter $w_i$
\end{quote}

In the case of the neuron model, 
\begin{align}
    \frac{\partial \mathcal{L}}{\partial \overline{g_{Na}}} = \frac{\partial \mathcal{L}}{\partial V} \frac{\partial V}{\partial \overline{g_{Na}}}
\end{align}

Importantly, both partial derivatives must be found in order to evaluate the gradient of the error with respect to the parameters, in this case $\overline{g_{Na}}$. 

The first term, $\frac{\partial \mathcal{L}}{\partial V}$, is the partial derivative of the objective function which is a function of $V$.  In neuron modeling, the choice of error function is up to the modeler. Evolutionary optimization methods often use multi-objective functions that aim to match the statistical features of model voltage traces to a target \citep{van_geit_bluepyopt_2016}. However many of these objective calculations are not feasibly differentiable.  Therefore, for this demonstration, we will use the Mean Squared Error (MSE) objective function: 
\begin{align}
    \mathcal{L} = \frac{1}{2}\sum_i^n (\hat{V} - V)^2
\end{align}
where
\begin{quote}
    $\hat{V}$ is the target output voltage trace
\end{quote}

The partial derivative of MSE with respect to V is: 
\begin{align}
    \frac{\partial \mathcal{L}}{\partial V} = - \frac{1}{n} \sum^n_i (\hat{V} - V)
\end{align}
Therefore the first term in backprop is differentiable. We will discuss how $V$ is acquired following our analysis of the second term in backprop.

The second term, $\frac{\partial V}{\partial \overline{g_{Na}}}$, is the partial derivative of the model solution with respect to the parameter $\overline{g_{Na}}$. It must then be possible for function $V$ to be differentiable with respect to $\overline{g_{Na}}$. To find the solution $V$, one must solve $\frac{dV}{dt}$, which is in part defined by $\overline{g_{Na}}$ (see equation~\ref{eq:hh}). Solving $\frac{dV}{dt}$ analytically is infeasible, so we must solve it numerically. To do in a way that enables the use of backprop, we must use an ODE solver that is fully differentiable. Using package torchdiffeq, we used a differentiable ODE solver implemented in PyTorch \citep{chen_neural_2019}. PyTorch takes advantage of the automatic differentiation programming library to guarantee that all operations used to numerically solve for $\frac{dV}{dt}$ to find $V$ are differentiable \citep{paszke_pytorch_2019}. We used relative standard parameters for the $ODESolve$, using the adjoint solver with method 'dopri5', with absolute and relative tolerances as `tol=1e-8 and rtol=e-8.

We can then describe this process with the following form: 
\begin{align}
    V = ODESolve(\frac{dV}{dt})
\end{align}
Therefore, to find the value of the second term, we must calculate: 
\begin{align}
    \frac{\partial V}{\partial \overline{g_{Na}}} = \frac{\partial}{\partial \overline{g_{Na}}} ODESolve(\frac{dV}{dt})
\end{align}

The auto-differentiation library makes this calculation feasible. Therefore, it is possible to calculate $\nabla \mathcal{L}$ with respect to each parameter for our multicompartment neuron model with active dendrites.


\section{Results}

\subsection{Fitting a heterogeneous active dendrite toy model using gradient descent}

\begin{figure}[!t]
  \centering
  \includegraphics[width=0.8\linewidth]{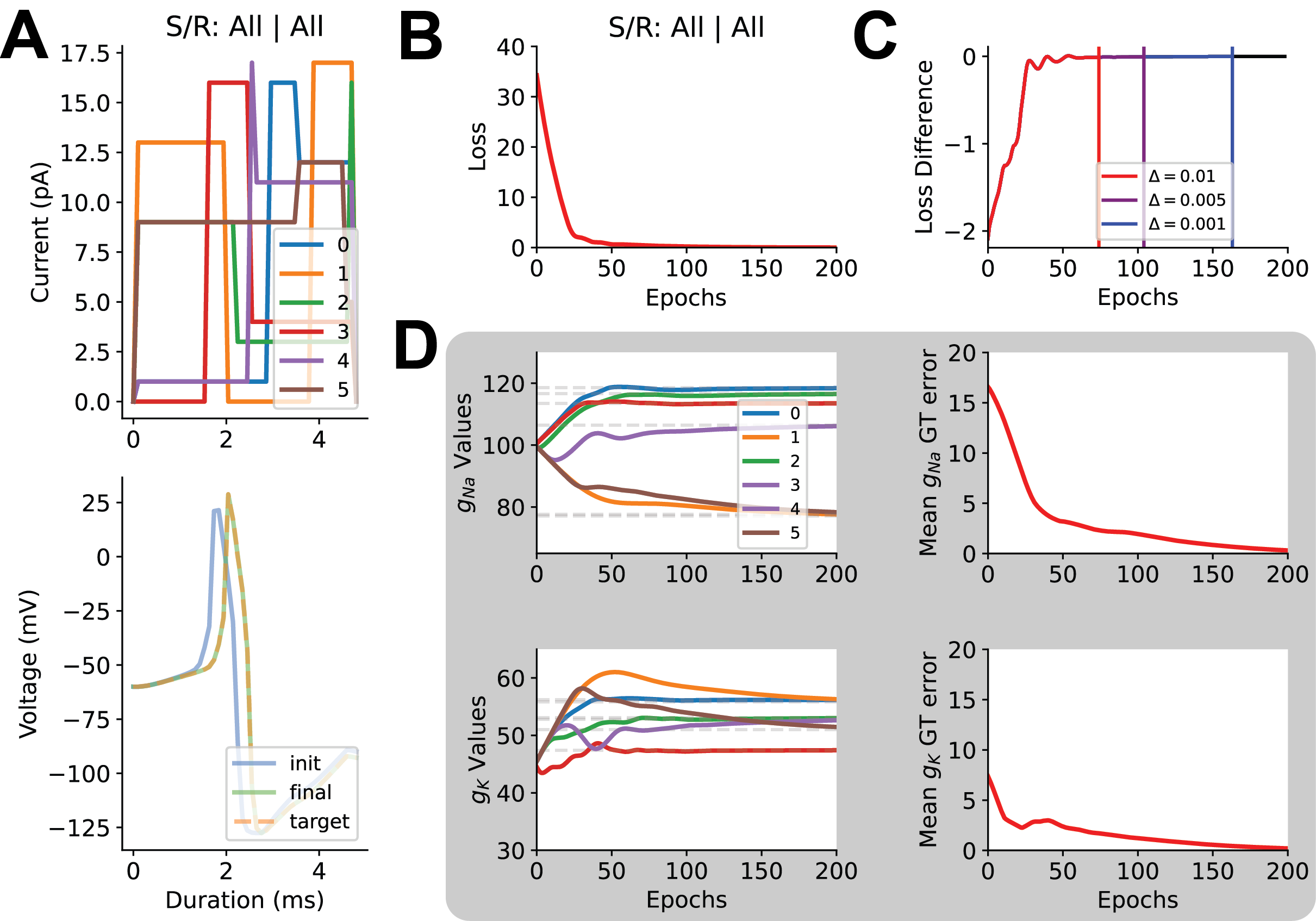}
  \caption{\textbf{Optimized Toy model successfully converges to matches target output and recovers ground truth parameters.} (a) Example input and resulting output before and after training. Top: Variable step current as stimulus to all compartments. Bottom: Somatic membrane voltage trace. Initial model output is in indigo. Target model output is dashed pink. Final model output is cyan. (b) Top: Loss over training converges and stops at epoch 200 with loss 0.006. Bottom: Trajectory of $\overline{g_{Na}}$ and $\overline{g_{K}}$ parameters for all compartments were initialized at 100 mS and 45 mS and ended near the randomized ground truth parameter values of the target model. (c) Points at which the loss would trigger a hypothetical early stopping condition. The difference in loss per epoch ($\Delta$) reaches the threshold at 74 ($\Delta = 0.01$), 104 ($\Delta = 0.005$), and 163 ($\Delta = 0.001$) epochs. Given an average epoch duration of 83.5 seconds, training would stop at 1hr 42min, 2hr 24min, and 3hr 47min respectively. (d) Distance from ground truth values decrease and mean GT errors converge to 0.318 and 0.207 for $\overline{g_{Na}}$ and $\overline{g_{K}}$ respectively.
}
  \label{fig:toy1}
\end{figure}

We will first use a simple toy model to check if our approach works. 
We thus introduce an idealized dataset where we expect gradient descent to minimize loss effectively. We first visualize the target trace and the initial and final training model traces (Figure~\ref{fig:toy1}A). We used a variable step current to stimulate all compartments of the toy model. To produce a simulated "ground truth" (GT) dataset, we randomized the parameter values and recorded the voltage trace outputs for all compartments. Figure~\ref{fig:toy1}A  shows an example current input to each of the compartments and the somatic target voltage trace $\hat{V}$ below as the dashed orange line. We then reinitialized all of the parameters of the training model to constant values $\overline{g_{Na}} = 100 mS$ and  $\overline{g_{K}} = 45 mS$. The resultant initial somatic voltage trace for the training model is in dark blue. We then used gradient descent using up to 200 iterations or epochs to attempt to minimize the loss. We computed the loss by comparing the resultant training voltage traces to the target voltage traces from all compartments. At the end of training, the final somatic voltage trace in green can be seen completely overlapping the target. With this simple model we can thus test the components of our approach.

We can visualize the the minimization of the loss metric over the process of gradient descent.  In Figure~\ref{fig:toy1}B, we show the trajectory of the normalized loss (loss divided by recorded model compartments), which has a final value of 0.001, a 98\% decrease from its initial value (Table~\ref{tab:table_rec_all}). The model was run for an arbitrary 200 epochs over the course of 4.5 hours (Table~\ref{tab:table_rec_all}). However if an early stopping criterion was implemented, training would end sooner (Fig.~\ref{fig:toy1}C). This demonstrates that gradient descent can reduce the loss in of the model output to the target output to close to zero.

It remains to be known if the model parameters found by gradient descent were in a local minima far from the ground truth parameter values. In Figure~\ref{fig:toy1}C we visualized if the parameter trajectories during training converged to the ground truth parameter values. We also introduced a Ground Truth Error metric (GT Error) to quantify the mean of the distances of all parameter values from the target GT parameter values. It is as follows:

\begin{align}
E_{gt} = f(g^* , g) = \frac{1}{N}\sum_i^N | g_i^* - g_i|
\end{align}

Where $g^*_i$ is the target parameter and $g_i$ is the model parameter. N is the number of compartments. 

It is clear in Figure~\ref{fig:toy1}C, the model parameters approach the ground truth parameter values by the end of optimization, with mean GT Error values close to 0 and a decrease of over 97\% (Table~\ref{tab:table_rec_all}). From this we demonstrate that under ideal data conditions where all compartments are stimulated and recorded from, gradient descent can not only fit the model to the target neural data outputs, but also approximate the ground truth parameter values of the target model.

\subsection{Fitting a toy model using semi-idealized data conditions}

\begin{figure}[!t]
  \centering
  \includegraphics[width=0.8\linewidth]{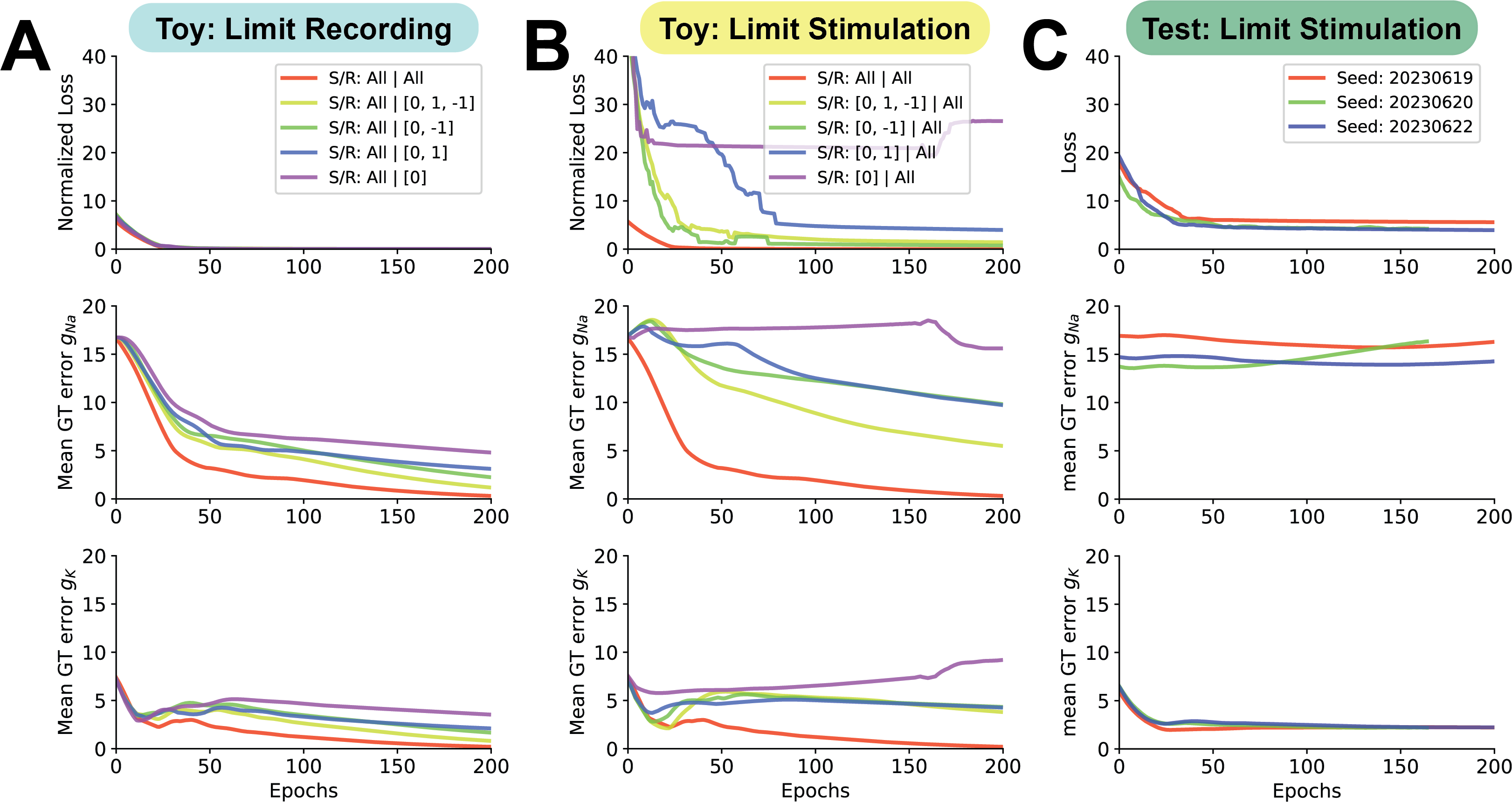}
  \caption{
  \textbf{Optimization metrics after limiting recording sites or stimulation sites in both Toy and Test models.} (a) In Toy model, Loss and Ground Truth Error after limiting recording sites and stimulating all compartments. Top: Loss over training converges for all conditions, even when recording sites are limited.  Bottom: Ground Truth Error for all parameters decreases consistently over training for $\overline{g_{Na}}$. Distance from ground truth for decreases non-monotonically over training for $\overline{g_{K}}$.  (b) In Toy model, Loss and Ground Truth Error after limiting stimulation sites and recording from all compartments. Top: Loss over training does not converge in most conditions when stimulation sites are limited, even when recording from all compartments. Bottom:  Ground Truth Error for all parameters decreases in some conditions, but does not approach zero by the end of training for both $\overline{g_{Na}}$ and $\overline{g_K}$. For conditions in legend, 'S/R' corresponds to 'Stimulation/Recording'. Conditions are '"All" : All compartments', '[0,1,-1] : Soma, Proximal and Distal Dendrites', '[0,-1] : Soma and Distal', '[0,1] : Soma and Proximal',  and '[0] : Soma only'. (c) In Test Model, Loss and Ground Truth Error of 3 seeds (see legend) after limiting stimulation sites and recording from all compartments. Top: Loss for all models converges and plateaus further away from zero. Bottom: Mean Ground Truth Error for $\overline{g_{Na}}$ does not decrease at all. Mean Ground Truth Error for $\overline{g_{K}}$ does decrease, but plateaus further from zero. 
}
  \label{fig:toylims}
\end{figure}

It is not technologically possible to acquire such complete data outside of  \textit{in silico} contexts. Therefore it is important to test if gradient descent can still operate under semi-ideal data conditions. We thus chose to limit stimulation sites and recording sites, while recording or stimulating all compartments respectively. In the toy model, the limited sites we stimulated or recorded from were the soma, the proximal dendrite adjacent to the soma, and the distal dendrite furthest from the soma (Figure~\ref{fig:model_comp}C). We find that when less rich data is available it becomes more difficult or impossible to match the target output or recover the ground truth physiological parameters.

\begin{table}[!t]
\centering
\footnotesize
    \begin{tabular}{|l|l|l|l|l|l|}
        \toprule
        \multicolumn{6}{c}{\textbf{Toy Model: Limit Recording Sites, Stimulate All Sites}} \\
        \midrule
        Condition S/R           &  All | All &  All | [0, 1, -1] &  All | [0, -1] &  All | [0, 1] &  All | [0] \\ \hline
        Seed                    &   20230619 &          20230619 &       20230619 &      20230619 &   20230619 \\ \hline
        Initial Normalized Loss &      5.709 &             7.132 &          7.215 &          6.82 &      6.674 \\
        Final Normalized Loss   &      0.001 &             0.003 &          0.006 &         0.002 &      0.011 \\
        Loss Decrease (\%)       &     99.983 &            99.964 &         99.913 &        99.967 &     99.842 \\ \hline
        Initial $\overline{g_{Na}}$ GTE        &      16.58 &            16.747 &         16.747 &        16.747 &     16.747 \\
        Final $\overline{g_{Na}}$ GTE          &      0.318 &             1.196 &          2.266 &         3.128 &      4.821 \\
        $\overline{g_{Na}}$ GTE Decrease (\%)   &     98.082 &             92.86 &         86.466 &        81.321 &      71.21 \\ \hline
        Initial $\overline{g_{K}}$ GTE         &      7.382 &             7.215 &          7.215 &         7.215 &      7.215 \\
        Final $\overline{g_{K}}$ GTE           &      0.207 &              0.82 &          1.681 &         2.113 &      3.546 \\
        $\overline{g_{K}}$ GTE Decrease (\%)    &     97.196 &            88.632 &         76.697 &         70.71 &     50.854 \\ \hline
        Optimization Time (H:M:S) &    4:38:21 &           4:36:26 &        4:39:56 &       4:41:33 &    4:40:23 \\
        \toprule
        \multicolumn{6}{c}{\textbf{Toy Model: Limit Stimulation Sites, Record All Sites}} \\ 
        \midrule
        Condition S/R           &  All | All &  [0, 1, -1] | All &  [0, -1] | All &  [0, 1] | All &  [0] | All \\ \hline
        Seed                    &   20230619 &          20230619 &       20230619 &      20230619 &   20230619 \\ \hline
        Initial Normalized Loss &      5.709 &            46.422 &         47.394 &         52.52 &     50.171 \\
        Final Normalized Loss   &      0.001 &             1.439 &          0.748 &          3.98 &      26.54 \\
        Loss Decrease (\%)       &     99.983 &            96.901 &         98.422 &        92.423 &     47.101 \\ \hline
        Initial $\overline{g_{Na}}$ GTE        &      16.58 &            16.914 &         16.914 &        16.914 &     16.747 \\
        Final $\overline{g_{Na}}$ GTE          &      0.318 &             5.513 &          9.844 &         9.748 &     15.605 \\
        $\overline{g_{Na}}$ GTE Decrease (\%)   &     98.082 &            67.407 &         41.799 &        42.366 &       6.82 \\ \hline
        Initial $\overline{g_{K}}$ GTE         &      7.382 &             7.215 &          7.215 &         7.215 &      7.548 \\
        Final $\overline{g_{K}}$ GTE           &      0.207 &             3.814 &          4.342 &         4.272 &      9.189 \\
        $\overline{g_{K}}$ GTE Decrease (\%)    &     97.196 &            47.138 &         39.825 &        40.795 &    -21.731 \\ \hline
        Optimization Time (H:M:S) &    4:38:21 &           4:18:47 &        4:57:49 &       8:23:16 &    7:58:22 \\
        \bottomrule
        
    \end{tabular}
    \caption{Optimization metrics (Loss and GT Error) for semi-idealized data conditions}
    \label{tab:table_rec_all}
\end{table}

We consider here the semi-ideal condition that all compartments of the toy model are stimulated and the number of recording sites are limited. All limited recording conditions approach a normalized loss close to 0 with decreases of over 99\% as seen in Figure~\ref{fig:toylims}A. All final model somatic voltage traces match the target voltage trace like that with the control (Fig.~\ref{fig:supptraces}A-E). All conditions begin approaching ground truth parameter values with similar trajectories, and it is possible that, with more training epochs, that each data condition could reach convergence to ground truth. Notably, even the soma-only recording condition could effectively fit the data with minimal normalized loss, as well as begin to approximate the GT parameters. Therefore, the semi-ideal condition where recording sites are limited but stimulation sites were comprehensive demonstrates success in approximating the input-output function and approximating the GT parameters.

We also consider the converse the semi-ideal condition that all compartments of the toy model are recorded but where the number of stimulation sites are limited.  The normalized loss metrics do not converge at 0 like the idealized condition control (Fig.~\ref{fig:toylims}B). This is supported in Figure~\ref{fig:supptraces}F-J, where the limited stimulation site conditions can be seen to show that some of the final fitted somatic voltage traces deviate from the target somatic voltage traces. Notably, the soma-only stimulation condition has the highest normalized loss relative to the other conditions with a decrease of only 47\%. (Fig.~\ref{fig:toylims}B). All the conditions including dendrite stimulation do not approach ground truth parameter values, but their GT Error trajectories do decrease, whereas the soma-only stimulation condition GT Error seems to stay the same or increase over training. Thus, the semi-ideal condition where stimulation sites are limited, demonstrates that this optimization method does not reliably fit the model to the target output or recover ground truth parameters, even when all compartments are recorded.

\subsection{Fitting a more realistic test model using gradient descent}

\begin{table}[!t]
\centering
\footnotesize
\begin{tabular}{|l|l|l|l|}
\toprule
\multicolumn{4}{c}{\textbf{Test Model: Limit Stimulation Sites, Record All Sites}} \\ 
\midrule
Seed                      &  20230619 &  20230620 &  20230622 \\ \hline
Initial Loss              &    18.052 &    14.812 &    19.174 \\
Final Loss                &       5.6 &     4.245 &     3.947 \\
Loss Decrease (\%)         &    68.977 &    71.343 &    79.413 \\ \hline
Initial $\overline{g_{Na}}$ GTE          &    16.914 &     13.73 &    14.715 \\
Final $\overline{g_{Na}}$ GTE            &    16.277 &    16.337 &     14.26 \\
$\overline{g_{Na}}$ GTE Decrease (\%)     &     3.763 &   -18.987 &     3.091 \\ \hline
Initial $\overline{g_{K}}$ GTE           &     5.975 &     6.501 &     6.434 \\
Final $\overline{g_{K}}$ GTE             &     2.225 &     2.203 &     2.233 \\
$\overline{g_{K}}$ GTE Decrease (\%)      &    62.755 &    66.107 &    65.302 \\ \hline
Optimization Time (H:M:S) &    3:1:47 &    3:1:47 &    3:8:34 \\
\bottomrule
\end{tabular}
\label{tab:table_test}
\caption{
Optimization metrics (Loss and GT Error) for Test model with realistic morphology.
}
\end{table}

It remains to be seen if a high-parameter model with realistic morphology could also be optimized to fit neural data. Here we introduce the test model, a branched 184 compartment model with 2 optimizable parameters each, yielding a 368 parameter model (Fig.~\ref{fig:model_comp}B). Like with the toy model, we randomized the parameters of the ground truth model. We then provided variable step current stimulation to 10 sites on the neuron model and recorded from all compartments. We optimized a new model initialized with constant $\overline{g_{Na}}$ and $\overline{g_{K}}$ parameters across the entire model and trained the model to attempt to minimize the loss between the output of the trained model output and the target model output. We repeated this process using 3 arbitrary seeds. In this setting, we sought to demonstrate that this test model can be trained to fit neural data using gradient descent.

With this semi-idealized data condition, we found that the loss trajectory over training decreased and converged to a value above 0, with a decrease of $\approx 68\%$. We repeated this training with 2 other models using other arbitrary seeds and received similar results (Fig.~\ref{fig:toylims}C). The GT Error did not converge for $\overline{g_{Na}}$, however some decrease occured for $\overline{g_{K}}$, though the values plateaued over training. This shows this high-parameter model likely converged to a local minima. The models took on average 3 hours to train, which is on the same order as the toy models. This demonstrates that gradient descent can scale well to neuron models with heterogeneous active dendrites resulting in high numbers of parameters.

\section{Benchmarking}

\begin{table}[!t]
\centering
\footnotesize
\begin{tabular}{|l|c|c|}
\toprule
& \textbf{\makecell{Evolutionary Algorithms \\ via DEAP in BluePyOpt}} & \textbf{\makecell{Gradient Descent \\ via Backprop in PyTorch}} \\
\midrule
\textbf{Compute} & 50 CPUs (Intel Xeon 2.60 GHz) & 1 GPU (Nvidia A100) \\ \hline
\textbf{Parameters} & 20 & \makecell{Toy: 12 \\ Test: 368} \\ \hline
\textbf{Duration (hours)} & 4 & \makecell{Toy: 4.5 (8) \\ Test: 3} \\
\bottomrule
\end{tabular}
\caption{Comparison of Evolutionary Algorithms and Gradient Descent}
\label{tab:benchmark}
\end{table}

We can compare gradient descent with backprop to BluePyOpt's evolutionary algorithm in 2016 \citep{van_geit_bluepyopt_2016} (Table~\ref{tab:benchmark}). The conditions where somatic stimulation only or combination of somatic and proximal dendrite stimulation had the longest optimization times. Otherwise, in cases with more stimulation the optimization time dropped to approximately 4.5 hours. It is worth noting that the Test model optimization times were shorter than the toy model. This may be a product of using smaller step currents in the Test model (Table~\ref{tab:simparams}) yielding faster forward simulation times in a stiff dynamical system. Gradient descent takes less computational resources while taking a similar or less amount of time to reach convergence. In comparison to BluePyOpt, gradient descent is more efficient and scaling to 20 times the parameters is feasible in compute time.

\section{Discussion}

In this paper, we have demonstrated that the backpropagation of error algorithm can fit the voltage output as a result of a given input in a biophysical conductance-based neuron model. We also demonstrated under ideal conditions of stimulating and recording all compartments, it can approximately solve the inverse problem and reveal the parameters of a ground-truth model. To do so it only required 3-4.5 or 8 hours using a single A100 GPU. When testing semi-ideal conditions to be loosely analogous to experimental conditions, we found that limiting the recording sites marginally reduced fitting performance, but limiting the stimulation sites greatly reduced fitting performance. We also demonstrated that this method can fit the parameters of a high-parameter, morphologically realistic neuron model with active heterogeneous dendrites without any significant increase in optimization time.

\subsection{Limitations and Possible Extensions}

Demonstrating that we can optimize the 6-compartment toy model and the 184-compartment test model demonstrates the versatility of gradient descent to optimize models with many parameters. However, neuron models have many current terms that rely on a variety of ion channel conductances. For example, BlyPyOpt fits up to 8 out of 10 different ion channel terms in their models using 0th order evolutionary algorithm methods per compartment type \citep{van_geit_bluepyopt_2016}. We fit only 2 different ion channel terms, which results in relatively few parameters given the complexity of neuronal physiology. If we were to reimplement the ion channels used in such models with a similar magnitude of compartments in our test model, such a heterogeneous active dendrite model would in theory have 4-5 times more parameters (totaling 1600-2000 parameters) and it is not clear how identifiable models with that many parameters would be. Further work can be done to implement models with more ion channel conductances which would demonstrate that gradient descent methods could realistically be used in neuron models with heterogeneous active dendrites.

In addition to the ion channel conductances that could be included in this model, we also focus on physiology in response to external input and do not include synapse properties or synaptic inputs. Pyramidal cells have on the order of 10s of thousands of synapses, a fraction of which can be active at any one time \citep{megias_total_2001, spruston_pyramidal_2008}. We do not include AMPA current inputs, which provide small and exponentially decaying currents into the dendritic trees \citep{spruston_pyramidal_2008, koch_biophysics_1999}, or NMDA current inputs, which can enable superlinear nonlinearities in the summation of local synaptic inputs in a dendritic tree or compartment \citep{schiller_nmda_2000, london_dendritic_2005, spruston_pyramidal_2008, tran-van-minh_contribution_2015}. Exploring the inferred patterns of synaptic weight distributions given designed or well-defined synaptic neurotransmitter inputs could allow for investigations how and why synapses may be clustered in dendritic trees. Following work by \cite{moldwin_perceptron_2020} using the perceptron learning rule, we lay the groundwork that argues it should now be possible to optimize synaptic weights in morphologically complex biophysical ODE neurons using 1st order gradient descent enabled by the backpropagation of error algorithm. 

Our demonstration of gradient descent to optimize ODE neuron models with heterogeneous active dendrites depends on short simulation durations of 5 ms, which is long enough to view exactly one spike. We made this decision for several reasons. One reason is that the time to complete optimization scales linearly with respect to the "walltime" it took to simulate the model. Focusing on one spike was sufficient for fitting the model we used when we provided stimulations to all compartments in the toy model, and effective for fitting the test model. However, it is possible that convergence could be faster with a simulation duration long enough for at least 2 spikes because it provides more data to fit with the MSE objective function. However, there could be a tradeoff between simulating more data for faster convergence, and lengthening walltimes for each epoch as convergence is reached. More fine-tuning of this process is possible to explore this tradeoff.

The second reason for using short simulation durations is that using MSE severely penalizes the false absence or presence of a spike. With long simulation durations on the order of 1 second, there could be many or no spikes in a given voltage trace output \citep{gouwens_systematic_2018}. This could potentially lead to poor convergence due to optimization to many local minima or saddle-points in the loss landscape. Fitting the shape of a single action potential is a more well-defined task for the MSE objective. Other 0th order algorithms such as evolutionary algorithms use multi-objective functions for features of voltage traces that are not limited to the shape or onset of one action potential, such as spike rate. However, we did not use these objectives due to the difficulty of implementing them as differentiable functions for use with backprop. Further work should be done to develop differentiable versions of alternative or multi-objective functions that scale better to feature statistics of voltage traces with longer simulation durations.

We simulated our idealized target data \textit{in silico} in a noiseless data regime. However real neural input and output physiology data contains noise from synaptic inputs and technical constraints with the physical medium of real neurons in slice physiology. Even though we were able to demonstrate that with the idealized data conditions that ion channel conductance parameters were identifiable using gradient descent, there are no guarantees that in a noisy data regime that this is possible. Additionally, the MSE objective function will also be sensitive to noise, potentially reducing its effectiveness in producing a model fit. Though we demonstrate the efficiency of 1st order optimization methods, this highlights the need of further work to test this method in a noisy data regime, and designing multiple objectives that may be less sensitive to noise when fitting neural data.

We used differentiable ODE solvers to numerically solve for voltage traces produced by biophysical ODE models. However, over the process of developing this method, several engineering-oriented adjustments needed to be made in order to avoid problems that arise from the sensitivities involved in solving ODEs numerically. One such example of this was adjusting the biophysical ODE terms avoiding certain state values from becoming zero due to bit estimation limits, which can result in operations involving division by zero, which are not differentiable. As such, this work is a proof of concept that 1st order methods enabled by differentiable ODEs is possible. Further work must be done to provide an accessible toolbox to the modeling community to work with simulated and real neural data.

\subsection{Contributions}
In this work we demonstrated that 1st order gradient descent optimization methods can feasibly optimize ODE neuron models with heterogeneous active dendritic compartments. We explain that the time complexity of 1st order methods scales linearly with respect to the parameters, which is one order more efficient than 0th order methods. This makes optimization of high-parameter models efficient and more feasible with less computational requirements. Optimization of one 368 parameter model using this method can be computed on 1 A100 GPU for 3 hours, as compared to optimization of one 20 parameter model for 4 hours using 50 CPU cores using evolutionary algorithms \citep{van_geit_bluepyopt_2016}. Recent work \citep{zhang_gpu-based_2022} involving the effectiveness of using GPUs to speed up the forward simulation of biophysical models promises to reduce our time simulation times further, which would linearly reduce our optimization times as well. Additionally, alternatives to PyTorch such as Jax and Diffrax \citep{bradbury_jax_2018, kidger_neural_2021} promises greater speed-up in the forward simulation. This work argues for the shift toward using these methods which are faster with less compute cost, which not only could allow for scaling up model fitting to multi-neuronal networks, but also decreasing the threshold for accessibility to the resources needed to fit neuron models (at the time of this writing, services such as Google Colab allows users to access a single A100 GPU, which was what was used to carry out all optimization in this paper.) Modelers may no longer need to have access to CPU compute clusters to fit neuron models.

We demonstrate in an idealized noiseless data regime that \textit{in silico} single neuron model parameters can be recovered using gradient descent. Testing neuron model optimization techniques to see recovery or approximation of ground truth parameters could be a means to test the quality of neuron model fitting methods. \cite{prinz_similar_2004} and other dynamical systems literature \citep{chis_relationship_2016, villaverde_structural_2016} has pointed out that many dynamical systems express degeneracy, leading to many possible sets of parameters that produce the same valid input-output relationship the model aims to be fit to. Model degeneracy is in part due to model bias (there may be too few parameters to fit the model), and the nature of the data in relation to the model. Lowering the barrier to optimizing neuron models may allow for easier exploration of theoretical data collection designs that suggests informative data perturbations that may help reduce model degeneracy. Testing multiple model possibilities \textit{in silico} that may be better specified to theoretical neural data would also be made more feasible as gradient descent allows for more degrees of freedom in model choice. Fast iteration enabled by fast optimization may make this process of approaching ground truth, or at least reducing model degeneracy, possible.

We tested semi-idealized neural data conditions where we limited recordings and limited stimulations to specific sites in our toy and test models. These semi-idealized conditions could loosely correspond to the type of data that could possibly be obtained using a combination of multi-site patch clamp, voltage imaging, and glutamate imaging or uncaging \citep{larkum_propagation_1996, aseyev_current_2023, aggarwal_glutamate_2023, ellis-davies_two-photon_2019}. This work suggests recording from all compartments of a branch of dendrite (toy model) or an entire dendritic tree (test model) with voltage imaging may have limited success if stimulation, or at least knowledge of more inputs, is not nearly as comprehensive. However, given that the optimization process had much more success in both minimizing loss and approximating ground truth parameters when all compartments were stimulated, this suggests this method has potential success with glutamate imaging techniques if the field of view is able to capture the full dendritic tree. The combination of gradient descent optimization on glutamate imaging data could potentially allow for  testing hypotheses involving ion channel conductance distributions across an active dendritic tree in slice physiology experiments. Investigating the smoothness of ion channel conductances distributed across the dendritic tree based on physiology is an example of the class of question that could be investigated that could aid in discussions of cell type taxonomy in health and disease.

We demonstrate 1st order gradient descent methods reduce time and computational resources needed to optimize biophysical neuron models with active heterogeneous dendrites. 0th order methods create a bottleneck in the number of parameters that can be possibly optimized in neuron models, for which there are efforts to effectively reduce complex biology of neurons so that model fitting is possible. Though these efforts enable asking questions with more interpretable simple models, the computational bottleneck of 0th order methods creates a bias in what kinds of questions are asked by modelers. Questions that were not feasible to answer because they required the use of high-parameter models may now be asked by neuroscientists with or without access to large compute clusters. However, the next step is for the development accessible tools for modelers to use 1st order gradient descent methods that can be added to current simulation environments such as NEURON \citep{hines_neuron_1997}. Such efforts could enable more standardized inquiries into the complexity of neurons at the neuronal level, involving intrinsic excitability of dendrites, patterning of synapses correlated in both their activity and spatial distribution in dendritic trees, and the testing of heterogeneous plasticity rules dependent on dendritic properties.

\section{Acknowledgments}

We would like to acknowledge the members of the Kording Lab, specifically Jordan Matelsky and Richard Lange, for help in the development of this project. We also acknowledge Carsen Stringer, Anton Arkhipov, Nathan Gouwens, and Kaspar Podgosky for their feedback during the development of this manuscript. We also would like to acknowledge Felipe Parodi and Joey Rudoler for their comments on the manuscript.  This work was funded by grants from the National Institute of Health, National Science Foundation, and Howard Hughes Medical Institute. This work has been made possible in part by a gift from the Chan Zuckerberg Initiative Foundation to establish the Kempner Institute for the Study of Natural and Artificial Intelligence.

\section{Code}
The code for this project can be found at \url{https://github.com/ilennaj/ode_neuron_backprop}. \citep{jones_ode_neuron_backprop_2024}

\bibliography{bibliography.bib}

\newpage

\appendix

\section*{Supplementary Figures}

\counterwithin{figure}{section}
\renewcommand{\thefigure}{S\arabic{figure}}
\setcounter{figure}{0}  
\begin{figure}[!ht]
  \centering
  \includegraphics[width=0.8\linewidth]{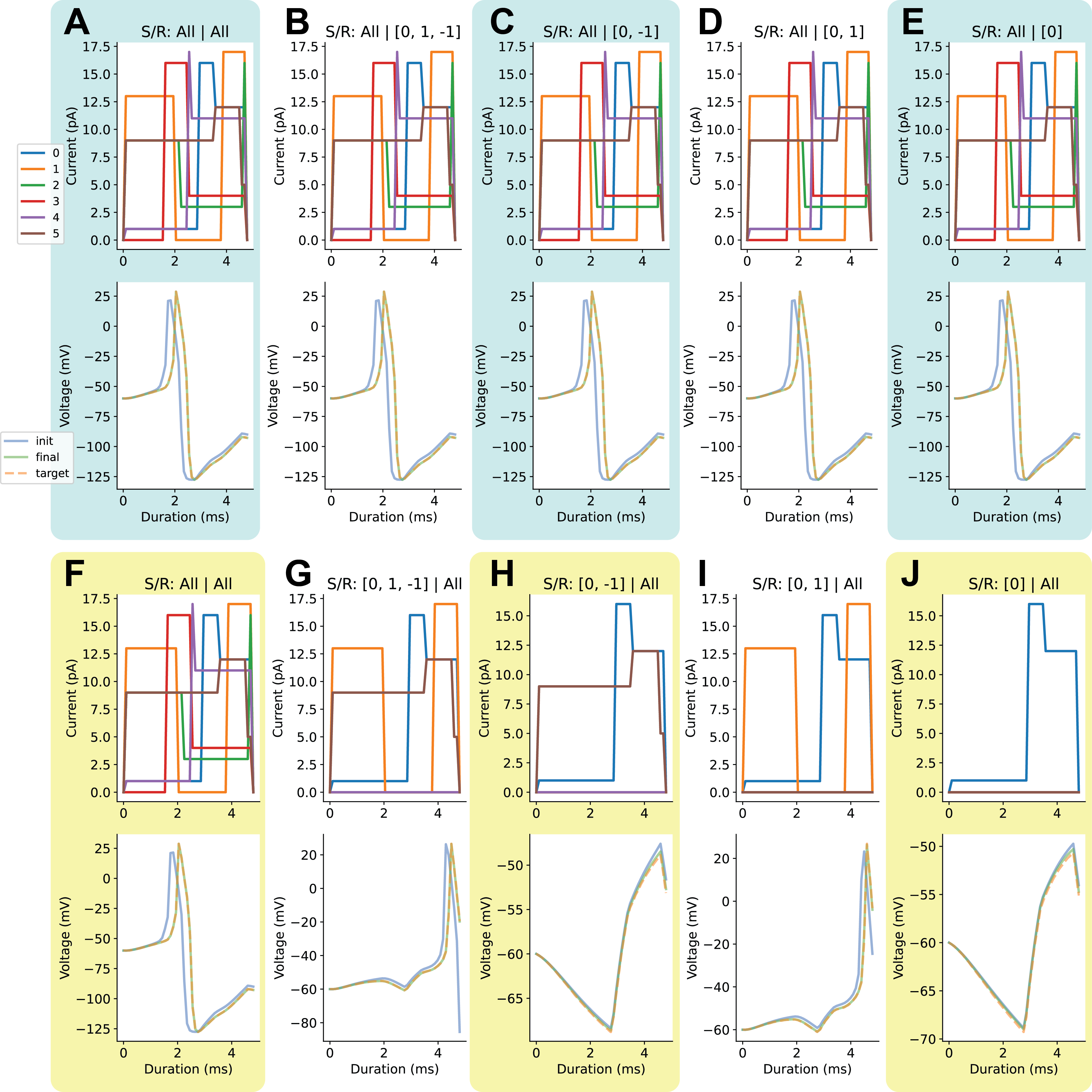}
  \caption{\textbf{Example voltage current and voltage traces for semi-idealized toy model conditions}. (a-e) Recording sites are limited and stimulation is given to all compartments. Top: Variable step current is given to all compartments. Bottom: Optimized model output matches the target output. Initial model output is in indigo. Target model output is dashed pink. Final model output is cyan. In all cases, the final model output matches the target. 'All | All' condition is present as a control for comparison. (f-j) Stimulus sites are limited and all compartments are recorded. Top: Variable step current is given to specific combinations of compartments. Bottom: Somatic voltage traces for each data limit condition before and after optimization. Not all conditions effectively match the final model output to the target output.
}
  \label{fig:supptraces}
\end{figure}

\end{document}